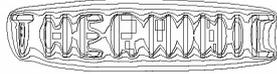

# QUANTITATIVE 3 OMEGA SCANNING THERMAL MICROSCOPY : MODELING THE TIP AC/DC COUPLING AND THE SAMPLE HEAT CONDUCTION

*Pierre-Olivier Chapuis, Sourabh Kumar Saha, and Sebastian Volz*

Laboratoire EM2C, CNRS and Ecole Centrale Paris,
Grande Voie des Vignes
92295 Châtenay-Malabry cedex, France

**ABSTRACT**

A way to increase the Scanning Thermal Microscope (SThM) sensitivity in the harmonic 3ω mode is to heat the probe with an AC current sufficiently high to generate a coupling between the AC and the DC signals. We detail in this paper how to properly take into account this coupling with a Wollaston-probe SThM. We also show how to link correctly the thermal conductivity to the thermal conductance measured by the SThM.

## 1. INTRODUCTION

The continuous decrease of sizes in the material science applications and in the microelectronics industry has led to the development of instruments that aim at analyze heat transfer at micro and submicroscales. Thermal characterization at nanoscales has recieved significant amount of efforts since a decade. While the optical methods have reached their maturity and remain limited by diffraction to half a micron, contact probes were considered as the only means to reach 10-100nm scale resolutions. The Scanning Thermal Microscope (SThM) based on an Atomic Force Microscope mounted with thermal probe was invented in 1986 by Williams and Wickramasinghe [1] to provide the topography of electrically insulating materials. Since then, various techniques have been proposed with spatial resolutions varying between 5 microns and a few tens of nanometers. However, quantitative measurements are still affected by two main artefacts, i.e. (i) the topography-thermal coupling and (ii) the thermal resistance between tip and sample. We previously showed that the artefact (i) is amplified by heat conduction through air due to poor spatial resolution [5]. The thermal resistance between the tip and the sample (ii) is due to conduction in air, through the solid-solid contact and through the water meniscus between tip and sample. The mechanical properties, the roughness and the hydrophilicity of the sample are involved in those interactions.

We will focus on the Wollaston wire probe [2-4]. A way to increase the SThM sensitivity in the harmonic (3ω) mode [5] is to heat the sample with an AC current sufficiently high to generate a coupling between the AC and the DC signals. We detail in this paper how to properly take into account this coupling. We also give the correct factor that takesnto account the constriction of the heat flux lines in the sample.

## 2. PROBE MODEL

The probe is a Wollaston wire consisting of a platinum core 5 microns in diameter and a silver coating 70 microns in diameter. The silver coating is etched to uncover the platinum wire over a $2L_s$=200 microns length. This tip was studied in several of our previous works [3-5]. A modulated (AC) electrical current is used to Joule heat the wire and the second harmonic of the temperature is measured with a Lock In Aplifier at the third harmonic of the voltage (see Reference [5]).

$$V_{3\omega} = \frac{R_o \alpha I_o \theta_{2\omega}}{2}$$

The previous papers did not take the coupling between the constant (DC) and the modulated (2ω) temperature into account. The AC-DC coupling is solved in this section when the tip is in contact and out of contact with the sample.

An estimation of the Biot number shows that heat conduction can be considered as monodimensional in the wire [4]. The size of the tip is micrometric, so we can use the classical Fourier conduction equation in the tip:

$$\frac{1}{a_s}\frac{\partial T}{\partial t} = \frac{\partial^2 T}{\partial x^2} - \frac{ph(T-T_o)}{\lambda S} + \frac{\rho I^2}{\lambda S^2}$$

where x denotes the abscissa in the probe (x=0 is at the silver-platinum interface and $x=L_s$ is the middle of the tip). $a_s$ and λ are the probe thermal diffusivity and conductivity. $S$ and $p$ correspond to the section and the perimeter of the wire probe. ρ is the tip electrical resistivity and h represents the heat transfer coefficient between the wire and the ambient. When the current is modulated:

$$I = I_o \cos(\omega t + \varphi_I)$$

Taking into account the dependance on temperature of the probe resistivity, it turns out that

$$\rho = \rho_o (1 + \alpha \delta T)$$

where α is the temperature coefficient of the wire.

The first equation can then be written as follows:

$$\frac{1}{a_s}\frac{\partial T}{\partial t} = \frac{\partial^2 T}{\partial x^2} - \frac{ph(T-T_o)}{\lambda S} + \frac{\rho_o I_o^2 (1+\alpha \delta T)(1+\cos(2\omega t))}{2\lambda S^2}$$

if we consider a time origin so that the phase is zero. We consider that the local temperature in the tip can be written as the sum of a DC term and a AC term:

$$\theta(x,t) = \theta_{DC}(x) + \theta_{2\omega}(x) e^{j2\omega t}$$

where

$$\theta = T - T_{amb}$$

Note that the AC term is complex : a phase lag between the current and the intensity is possible. Thus we can decompose the equation for $\theta$ into its DC component:

$$\frac{d^2 \theta_{DC}}{dx^2} - \frac{ph\theta_{DC}}{\lambda S} + \frac{\rho_o I_o^2}{2\lambda S^2} + \frac{\rho_o \alpha I_o^2 \theta_{DC}}{2\lambda S^2} = 0$$

and into its 2ω harmonic component:

$$j\frac{2\omega}{a}\theta_{2\omega} = \frac{d^2 \theta_{2\omega}}{dx^2} - \frac{ph\theta_{2\omega}}{\lambda S} + \frac{\rho_o I_o^2}{2\lambda S^2}(\alpha \theta_{2\omega} + \alpha \theta_{DC} + 1)$$

The DC and AC equations are coupled through the dependence of the harmonic temperature on the DC temperature. The here-called DC equation is the same as the one obtained in the DC regime if the Joule term is divided by a factor of 2.

The general solution of the DC equation is

$$\theta_{DC}(x) = c_1 e^{m_1 x} + c_2 e^{-m_1 x} + \frac{J}{m_1^2}$$

where

$$m_1^2 = \frac{ph}{\lambda S} - \frac{\rho_o I_o^2 \alpha}{2\lambda S^2}$$

and

$$J = \frac{\rho_o I_o^2}{2\lambda S^2}$$

The general solution of the AC equation is:

$$\theta_{2\omega} = c_3 e^{m_2 x} + c_4 e^{-m_2 x} + A e^{m_1 x} + B e^{-m_1 x} + C$$

where

$$m_2^2 = \frac{ph}{\lambda S} - \frac{\rho_o I_o^2 \alpha}{2\lambda S^2} + j\frac{2\omega}{a_s}$$

and

$$A = -\frac{\alpha J c_1}{m_1^2 - m_2^2}$$

$$B = -\frac{\alpha J c_2}{m_1^2 - m_2^2}$$

$$C = \frac{J(1+\alpha \frac{J}{m_1^2})}{m_2^2}$$

The constants $c_1, c_2, c_3$ and $c_4$ are obtained from the boundary conditions. The symmetry of the problem allows us to solve the problem for only one half of the probe tip. The silver end of the probe is a thermal sink for the harmonic component. For the DC component, the flux at the silver end is taken into account through the high conductance of one of the legs of the silver coating [4]. The other end can be considered insulated due to symmetry when the tip is out of contact with the sample. The boundary conditions are described as follows.

At x=0:

$$\lambda S \frac{d\theta_{DC}}{dx} = G_{Ag}\theta_{DC}, \theta_{2\omega} = 0$$

At x=$L_s$:

$$\frac{d\theta_{DC}}{dx} = 0, \frac{d\theta_{2\omega}}{dx} = 0$$

When the tip is in contact, the boundary conditions are modified by the heat flux from the tip to the sample. We assume that this flux is confined in the volume around the tip/sample contact. The heat transfer through air is neglected on the remaining surface of the tip. We argue that the hotter part of the tip is in the vicinity of the sample. This was confirmed by simulations in a previous paper [6]. When the pressure decreases, the mean free path of the air molecules increases, for instance $10^{-2}$ bar corresponds to a 10 microns mean free path. But the surface of the tip legs have a lower contribution because

(i) the temperature is lower on these legs than in the middle of the tip and (ii) also note that the molecule-molecule scattering is still lower in the heat transfer between the very end of the tip and the sample than the one between the tip legs and the contact area. We finally write at x=Ls:

$$-\lambda S \frac{d\theta_{DC}}{dx} = G'_{eq,DC}\theta_{DC}, \quad -\lambda S \frac{d\theta_{2\omega}}{dx} = G'_{eq,AC}\theta_{2\omega}$$

where S is the contact surface and the prime on $G_{eq}$ denotes that it is half of the total conductance since the equations are solved only for one half of the tip. We surely do not reproduce the real heat flux lines in the tip-sample contact area. But the size of this area is small in comparison to the one of the tip. The average temperature will not be significantly modified.

Note also that $G_{eq,AC}$ is not known yet. We assume that $G_{eq,AC}=G_{eq,DC}=G_{eq}$ because the thermal characteristic time is smaller than the reverse of the frequency $f=3\omega/2\pi$. For heat transfer through air, the characteristic time is the diffusion time through air over 100 microns at atmospheric pressure, which is of the order of $10^{-4}$ s. The heat transfer through the meniscus and the mechanical contact are faster because the system sizes are of the order of nanometers. This justifies the equivalence of the tip-sample heat transfers in the DC mode and in the AC mode in our range of frequencies.

Applying the boundary conditions we have:

$$c_1 = \frac{-J}{2m_1^2}\left[\frac{\left(G_{Ag}+\lambda S m_1 - G_{Ag}e^{-m_1 L}\right)G'_{eq}+G_{Ag}\lambda S m_1 e^{-m_1 L}}{\lambda S m_1(G'_{eq}+G_{Ag})\cosh(m_1 L)+\left\{(\lambda S m_1)^2+G'_{eq}G_{Ag}\right\}\sinh(m_1 L)}\right]$$

$$c_2 = \frac{-J}{2m_1^2}\left[\frac{(-G_{Ag}+\lambda S m_1+G_{Ag}e^{m_1 L})G'_{eq}+G_{Ag}\lambda S m_1 e^{m_1 L}}{\lambda S m_1(G'_{eq}+G_{Ag})\cosh(m_1 L)+\left\{(\lambda S m_1)^2+G'_{eq}G_{Ag}\right\}\sinh(m_1 L)}\right]$$

and

$$c_3 = -\left[\frac{A(G'_{eq}+\lambda S m_1)e^{m_1 L}+B(G'_{eq}-\lambda S m_1)e^{-m_1 L}-(A+B+C)(G'_{eq}-\lambda S m_2)e^{-m_2 L}+G'_{eq}C}{2(\lambda S m_2 \cosh(m_2 L)+G'_{eq}\sinh(m_2 L))}\right]$$

$$c_4 = \left[\frac{A(G'_{eq}+\lambda S m_1)e^{m_1 L}+B(G'_{eq}-\lambda S m_1)e^{-m_1 L}-(A+B+C)(G'_{eq}+\lambda S m_2)e^{m_2 L}+G'_{eq}C}{2(\lambda S m_2 \cosh(m_2 L)+G'_{eq}\sinh(m_2 L))}\right]$$

with $G'_{eq}=0$ for the out of contact case.

At this step, we know $\theta(x)$. What is measured in the 3ω mode is the mean temperature of the tip. It is found by integration over the length of the tip.

## 3. RESULTS

We present on the following figure the results showing that the coupling is responsible for the heating of the probe of a temperature on the order of 5 K for an input current of 50 mA at ambient pressure (here h=3500 Wm$^{-1}$K$^{-1}$). If the coupling was not taken into account, a bad exploitation of the results could lead to an error of one order of magnitude in the conductance calculation. One should note as seen on Figure 2 that the coupling does not affect the phase lag : the two curves (with and without coupling) are completely superimposed.

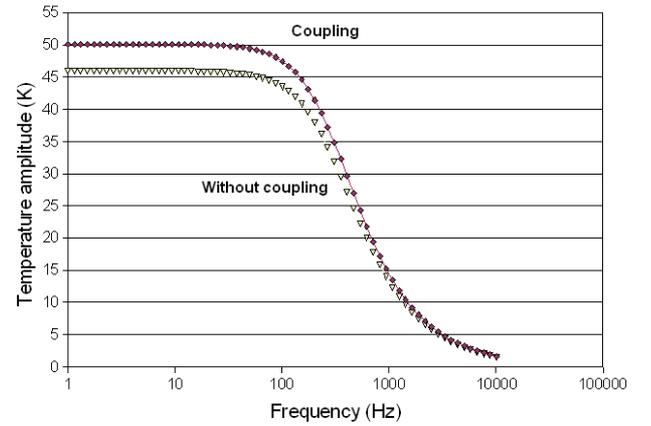

*Figure 1. Harmonic temperature amplitude calculated as a function of the frequency of the input current : cases with and without AC/DC coupling.*

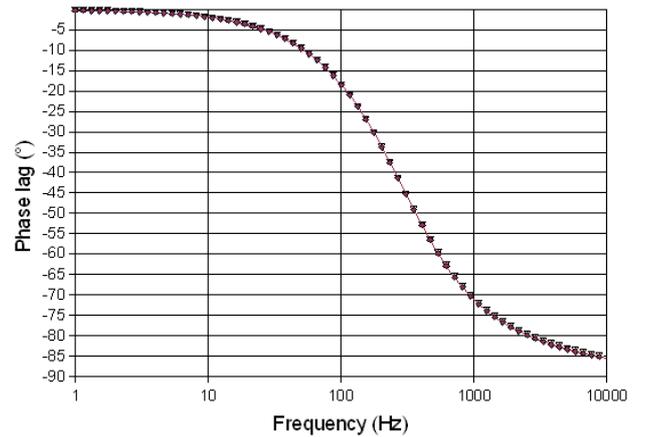

*Figure 2. Phase lag of the harmonic temperature calculated as a function of the frequency of the input current : there is no difference with or without coupling*

## 4. HEAT FLUX IN THE SAMPLE

We have considered in the first part the heat flux at the contact by using a the thermal conductance $G_{eq}$. Its physical meaning corresponds to the contributions of the contact, $G_c$, and the sample, $G_s$, conductances:

$$\frac{1}{G_{eq}} = \frac{1}{G_S} + \frac{1}{G_C}$$

The contact conductance $G_c$ takes into account conduction in air ($G_a$) and in the water meniscus ($G_w$) and through the solid-solid mechanical contact ($G_{ss}$). We stress in the following that a summation rule is applicable

$$G_c = G_a + G_w + G_{ss}$$

The sample conductance $G_s$ corresponds to the effect of constriction of the heat flux lines in the semi-infinite sample.

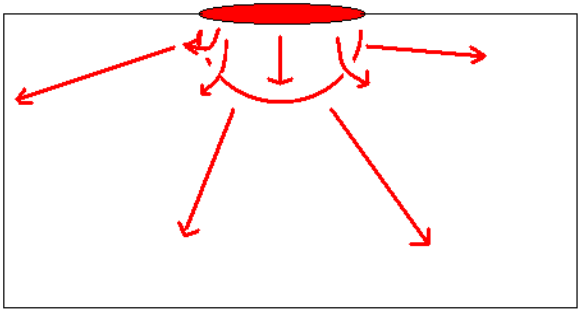

*Figure 3. Heat flux lines arising from the hot disk of radius b on the surface.*

The heat flux in the sample can be described by the product of the temperature on the sample surface $\theta_s$, the contact radius *b* and the sample thermal conductivity $\lambda_s$:

$$\phi \sim \lambda_s \, b \, \theta_s$$

If we consider that the heat flux generated by the tip in the sample arises from a hot disk on the top of the sample (see Figure 3), the geometrical factor taking into account the constriction of the heat flux lines is $3\pi^2/8$ and we have then :

$$\phi = \frac{3\pi^2}{8} \lambda_s b \theta_s$$

This relation is retrieved by performing a finite element simulation of a hot disk on the top of the semi-infinite volume. The geometrical factor which is independent of the thermal conductivity of the material matches with the analytical one within 10%. Note that this is the first time that this geometrical factor is correctly taken into account : there was previously [3-5] a confusion with the geometrical factor of the lines arising from a semi-hemisphere ($2\pi$).

## 5. CONCLUSIONS

We presented here a new step on the road to quantitative Scanning Thermal Microscopy by showing how to take into account the AC/DC coupling at high input currents, which are necessary if one wants to increase the sensitivity of the instrument. We also showed how to link correctly the thermal conductivity and the heat flux flowing from the tip to the sample by introducing the correct geometrical factor.